\documentclass[elsart,elsart12,epsfig,epsf,graphics,longtable,amstex,amssymb,euscript]{elsart}
\usepackage{epsfig}
\usepackage{amssymb}

\begin{document}
\begin{frontmatter}
\title{Novel Technique for Ultra-sensitive Determination of
Trace Elements in Organic Scintillators}
\author{Z. Djurcic$^1$}
\author{D. Glasgow$^2$}
\author{L-W. Hu$^3$}
\author{R.D. McKeown$^4$}
\author{A. Piepke$^1$}
\author{R. Swinney$^4$}
\author{B. Tipton$^4$}
\address{\small \it $^1$ Department of Physics and Astronomy, University of Alabama, Tuscaloosa, AL}
\address{\small \it $^2$ Oak Ridge National Laboratory, Oak Ridge, TN}
\address{\small \it $^3$ Nuclear Reactor Laboratory, MIT, Cambridge, MA}
\address{\small \it $^4$ W.K. Kellogg Radiation Laboratory, Caltech, Pasadena, CA}
\maketitle
\begin{abstract}

 A technique based on neutron activation has been developed
for an extremely high sensitivity analysis of trace elements in organic
materials.  Organic materials are sealed in plastic or high purity quartz
and irradiated at the HFIR and MITR.
The most volatile materials such as liquid scintillator (LS) are first 
preconcentrated
by clean vacuum evaporation.  Activities of interest are separated
from side activities by acid digestion and ion exchange.  
The technique has been applied to study the liquid
scintillator used in the KamLAND neutrino experiment.  
Detection limits of $<$2.4$\times$10$^{-15}$ g $^{40}$K/g LS, 
$<$5.5$\times$10$^{-15}$ g Th/g LS, and $<$8$\times$10$^{-15}$ g U/g
LS have been achieved.
\end{abstract}
\end{frontmatter}

\section{Introduction}

 A next generation of nuclear and particle physics
experiments will require increasingly pure materials for success.
The signatures of extremely rare processes such as neutrino
interaction and oscillation, double beta decay, and dark matter
interaction, are often masked by common terrestrial backgrounds.
New underground facilities may shield experiments from the
background derived from cosmic radiation. However, isolating the
physics processes from radiation
in the experimental
 apparatus itself remains critical.  Radionuclides such as
 uranium, thorium and potassium-40 have long half
lives and  remain abundant in the earth's crust.  The decay products of
 these nuclei and their daughters can often have the same low
energy signatures of the rare processes studied.
     To enable experiments probing rare phenomena, new research
into the selection of materials with very high purity in radionuclides
is key.

\section{KamLAND physics and requirements}

A prime example is the KamLAND experiment
~\cite{kam_prop}, which motivated this work.
KamLAND, installed in the cavity of the concluded Kamiokande experiment 
in the Japanese Kamioka underground laboratory, 
is detecting neutrino oscillations in disappearance mode
using Japanese nuclear power reactors as anti-neutrino sources ~\cite{prl_first}.
Its long baseline gives KamLAND sensitivity to 
$\Delta m^2$-values
in the range of the so-called large mixing angle solution of the
solar neutrino problem. The flux weighted mean distance to
22 nuclear power stations in Japan and Korea is about 200 km. 
This large distance to the neutrino
sources, which is essential for the physics, 
results in an extremely small anti-neutrino interaction 
rate of only about
2 per day in KamLAND's 1000 tons fiducial volume,
consisting of organic liquid scintillator.\\ 
Owing to the low
anti-neutrino energies, physics events can be masked by natural,
anthropogenic and cosmogenic radioactivity. 
It has been estimated, that 
the concentrations of $^{40}$K, $^{232}$Th and $^{238}$U should not exceed
$10^{-14}$ g/g in order to not compromise the physics measurement~\cite{kam_prop}. 
As an example of the requirements in another organic material, concentrations
 of Th and U in the acrylic vessel of the Sudbury Neutrino Observatory (SNO)
 were measured to be less than the specified limits of 10$^{-13}$ g/g
 ~\cite{SNO}.
The goal
of this work was therefore to test samples of the scintillation fluid for the 
presence of above elements
at these very small concentrations.\\

\section{NAA motivation}

While direct radiation counting is the ideal analysis technique for radiation
sensitive experiments, such analysis suffers
from a lack of decay rate. As an example, at a concentration of $10^{-14}$ g/g
U the expected decay rate is about 1 per day per 100 kg of sample. The
large sample
size needed therefore becomes impractical.
At the same concentration a 1 kg sample contains about $2.5\times 10^{10}$ uranium atoms.
 Direct detection of this large population of parent nuclei can be
 achieved with 
inductively coupled plasma mass spectroscopy (ICPMS) and neutron
activation analysis (NAA).  The secular equilibrium assumption in the uranium and thorium decay
chains relate the parent population measurements to the expected radiation.

Using efficient ion sources, sensitivities to $^{238}$U of the required magnitude
or even better have been reported in the literature~\cite{wasserburg}. 
However, these methods require
the nuclides of interest to be in aqueous solution. In order to
apply mass spectroscopy to organic liquids requires the blank free, efficient
transfer of the elements of interest into aqueous solution. 
This in turn requires chemical manipulations before the analysis.
The purity of the used chemicals is hence crucial for achieving a low
blank reading.
The basic principles governing trace element analysis by means of NAA
have been worked out in great detail in the past.
They shall hence not be discussed in this article.
Textbooks such as reference ~\cite{naa_book} give a comprehensive overview.
A further important criterion which led us to choose NAA as analysis tool
was the fact that sensitivities even exceeding those required for our work
have been reported in the literature~\cite{munich_borexino}, proving 
feasibility.
Organic substances are, at least in principle, well suited for NAA. 
Hydro-carbons, consisting of C, H, N, O have no long lived activation products.
The nuclides of interest, $^{41}$K, $^{232}$Th and $^{238}$U on the other
hand, offer sizable capture cross sections and their activation products,
$^{42}$K, $^{233}$Pa and $^{239}$Np, have half lives greater than 12
hours, stable
enough to allow decay radiation counting
even after substantial shipping and processing delays.\\ 
A problem, however, arises from the fact that hydrocarbons tend to dissociate 
in a high neutron and gamma flux environment. This so called radiolysis leads to
hydrogen outgassing  inside the nuclear reactor used for the
activation. In the previous work by the Munich group~\cite{munich_borexino}
pressure build-up in the sample containment vessel (and its potential
rupture associated with sample loss) was prevented by venting the
excess gas into the pool water of the reactor~\cite{kim}.
This approach has been used for liquid scintillator samples with up
to 500 ml volume at the research reactor, FRM I, in Munich~\cite{munich_98}.
Detection limits of $3\times 10^{-12}$ g/g for K, $2\times 10^{-15}$ g/g 
for Th and $2\times 10^{-16}$ g/g for U have been reported in the 
literature~\cite{munich_borexino}.  More recently, the Munich group obtained
further improved sensitivities of 1.8$\times$10$^{-16}$ g/g for 
Th and 1.0$\times$10$^{-17}$ g/g for U~\cite{munich_2}.
The direct irradiation of liquid samples 
does not require extensive sample preparation or treatment prior to the analysis,
enhancing the robustness against the introduction of unwanted contaminations.
Although this procedure offers a number of positive features its use seems to be
limited to the research reactor facility in Munich. No research reactor could be found
in the US which has an existing facility that would allow the irradiation
of large liquid samples and the venting of potentially flammable gases into
the coolant.

We therefore decided
to develop a novel technique, not requiring the venting of excess
gases.
In the approach described here, the organic materials are gently
heated in a clean vacuum chamber, in order to evaporate most of
the organic until less than one percent by mass remains.  Sample outgassing is 
substantially reduced, while most of the
impurities survive evaporation and are analyzed in the residual organic.  
The sensitivities achieved with this new technique are not quite as stringent
as those reported by the Munich group but, to the best of our knowledge,
they are only exceeded by this group. The main advantage of the approach
described in this article 
lies in the fact that it can be used at almost any research reactor,
it does not require expensive, unique irradiation facilities, 
and it is therefore generally
applicable. Irradiation facilities with very high neutron flux may
also be used.
At FRM I the thermal neutron flux was limited to 
$2\times 10^{12}-1.3\times 10^{13}$ cm$^{-2}$s$^{-1}$. By comparison the thermal neutron 
flux available at the High Flux Isotope Reactor (HFIR) of Oak Ridge National Lab. and at the
Massachusetts Institute of Technology Research Reactor (MITR) of
MIT Nuclear Reactor Laboratory
is with $4\times 10^{14}$  cm$^{-2}$s$^{-1}$ and $5\times 10^{13}$  cm$^{-2}$s$^{-1}$, respectively,
substantially higher. Both facilities were utilized in this work.\\

\section{Procedure}
Our work focused on the analysis of radio impurities in KamLAND liquid 
scintillator and its components.
The scintillator is composed of 80\% dodecane (C$_{12}$H$_{26}$) and 20\% 1,2,4-trimethylbenzene
(pseudocumene, C$_9$H$_{12}$). 1.5 g/l solid 2,5-diphenyloxazole (PPO, C$_{15}$H$_{11}$NO),
acting as the primary fluor, is dissolved in it.

\subsection{Sample Preparation}
   The analyzed samples of liquid scintillator were obtained directly
from the KamLAND experimental site in Japan.  For sample
collection and storage, one liter PFA Teflon
containers were etched in sub-boiling 25\% ultra-pure nitric acid
and 25\% ultra-pure hydrochloric acid over a period of one week.
The containers were then rinsed with ultra-pure water (MilliQ)
and dried.  Liquid scintillator was accessed in a clean room surrounding
the liquid plumbing of the detector.   An access tap was cleaned
with alcohol and water, and then 5 liters of scintillator was
drained through it.  Scintillator was then emptied into the collection
containers, shaken and discarded.  A second filling with scintillator
was kept for analysis.
The Teflon containers were closed, heat sealed in class 100 grade
nylon bags, and shipped to the analysis
laboratories at Caltech and the University of Alabama in the United States.  

Samples required further preparation at the United States laboratories
prior
to neutron activation.
Evaporation of the liquid at low temperature
and under vacuum leaves behind most of the solid PPO powder, the least
volatile component. The PPO can therefore be used as
a convenient collection medium for the elements of interest. Since for
K, Th and U only very few volatile chemical compounds exist, we would expect that 
recovery for the elements of interest after evaporating the liquid should be almost 
quantitative. 

After evaporation the PPO was heated and transferred to a small
container for irradiation.  A plastic nalgene bottle of 1 cm diameter
and 2 cm length was used initially for this purpose;  the lid
was melted and fused to the bottle with a clean soldering iron.
To reach lower blank levels and allow for longer irradiations, 
high purity, Suprasil T21 quartz vessels 
from Heraeus Amersil were alternatively used in later irradiations.
  The quartz bottle
derived from 8$\times$10 mm quartz tube stock. A 10 cm length
was cut from the tube and blown shut in one end.  The center of
the tube was then pinched inwards until a 3 mm opening remained.  The
bottle was etched with a regimen of concentrated acids, 
followed by a rinse of ultrapure water.
For PPO transfer, the quartz was heated to near 80 $^{o}$C to ensure
efficient movement of the organic into the bottom of the bottle.
On several runs, the evaporation beaker was rinsed by an additional
one milliliter of liquid scintillator which was then also
poured into the quartz vessel, which was then dried in the same apparatus.
After loading the sample, the quartz vessels were sealed by melting 
the pinched area closed.  Liquid nitrogen was applied continuously
to the bottom of the bottle in order to freeze the sample, and prevent
mass loss through PPO melting and boiling\footnote{It is advisable
to perform this operation under vacuum to prevent potentially hazardous 
condensation of O$_2$}.  

The sample preparation and analysis strategy depends critically 
on the retention of the elements of interest
in the PPO. Our measurements of the retention efficiency are
discussed in section 5.

\begin{figure}[hhh!!!]
\begin{center}
\leavevmode
\mbox{\psfig{file=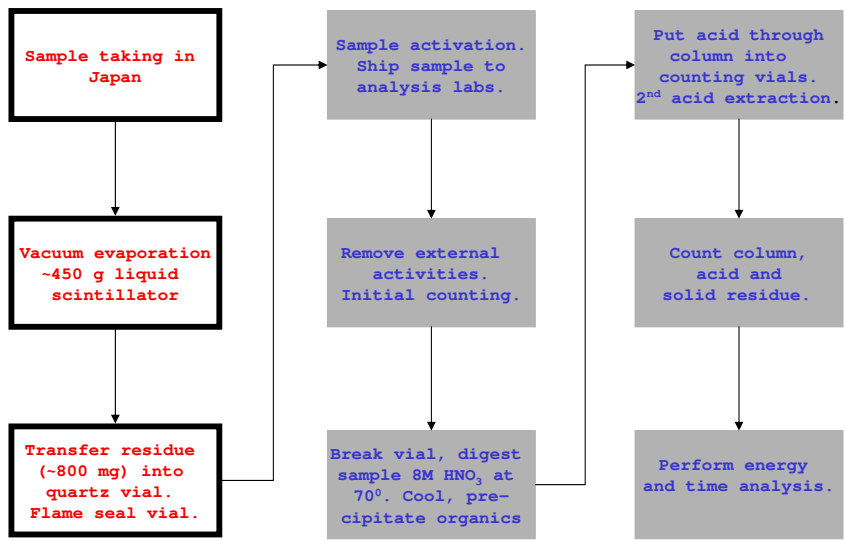,width=14cm,clip=}}
\end{center}
\caption{Flow diagram of the principal steps involved with the described analysis.
Steps displayed in a white box were performed under clean conditions
to prevent introduction of contamination. Once the samples were
activated no such precautions were needed.
The steps involving work with relatively high activity samples are displayed in grey boxes.}
\label{flow_fig}
\end{figure}

\subsection{Irradiation and Analysis}
Evaporated PPO samples were then irradiated.  A test run at HFIR in
October 2000 verified the feasibility of the techniques described
above.  The data reported here derive from irradiations at
MITR between January, 2001 and July, 2002.  
When encased in plastic bottles, the PPO samples
were shown to survive 15 minutes in HFIR or two hours in MITR
before sample boiling caused plastic bottle breaches and losses.  When
sealed in quartz, the sample containment survived at least 28 
hours of irradiation in the MITR 2PH1 pneumatic facility, 
without evidence of loss.  The neutron flux
was calibrated at MITR by irradiating trace amounts of salts of 
sodium, potassium, chromium, zinc, bromine, strontium, tin, antimony,
lanthanum, gold, thorium and uranium, all in identical quartz and 
plastic bottles.
The calibrated neutron flux 
varied from 3.0 to 4.3$\times 10^{13}$ thermal n/(cm$^2$ s)
and from 4.1 to 8.9$\times 10^{11}$ epithermal n/(cm$^2$ s) (neutron
energies larger than 0.5 eV), with
a 8\% precision in the determinations.

The pneumatic tube 2PH1 is one of the highest neutron flux irradiation
facilities at the MITR. This facility allows a sample to be inserted
pneumatically into a heavy water reflector tank reentrant thimble
for irradiation. Once the irradiation is terminated, the sample is ejected to
a hot cell that is used for temporary storage. A CO$_2$ purge is
maintained in the thimble which helps to reduce the amount of
$^{41}$Ar produced. The sample holder (rabbit) is a polyethylene
cylinder that has a 1-3/8'' ID and is 6-1/4'' long. Multiple PPO
samples can be accommodated in one rabbit.
    After irradiation samples were allowed to cool at a hot cell
for at least 12 hours, in order to minimize the radiation hazard of short-lived
activities, such as $^{31}$Si.  The samples were then packed for overnight
FedEX shipment to Caltech and UA. 
   The delay between sample ejection
from the reactor and receipt at the laboratories was reduced
to 35 hours, improving sensitivity to short-lived activities
such as $^{42}$K(T$_{1/2}$=12.36 hours).

  Upon receipt at the analysis laboratories, each sample 
was processed for radiation analysis
in a fume hood.  The processing involved several steps.  In the first step,
the exterior of each sample container, whether quartz or plastic, was 
cleaned, in order to remove 
external contamination unconnected with the sample.
  An initial acetone rinse removed residual organics and
connected activities.  Then each sample container was immersed in a small
 bottle containing 60 ml of concentrated nitric acid or aqua regia.
  After one hour, 
each sample was
moved to a fresh acid vessel and the process was repeated.  The acetone
and two acid washes were directly assayed to confirm that extractable
contamination sharply dropped from one wash to the next.  Studies with
additional 
exterior washes yielded external activities significantly 
lower than the activity extracted
from the interior of all samples, indicating the procedure was
effective.  Typical external activity reductions of $^{24}$Na greater
than 1000 were achieved. 
 No evidence of external uranium or thorium contamination
was seen in the washes, and only trace potassium levels were seen.

After exterior washes, the sample container was rinsed
by pure water, and then broken to expose its interior.  
The open containers were placed in beakers
to which strong nitric acid was added, in order
to extract the radioactivity from the sample itself.
The 
acid was heated to 80 $^{o}$ C on a hot plate for over a half hour.  At this
temperature, the sample of irradiated PPO was fully digested by the acid,
forming a transparent yellow solution.  The sample container was extracted from
the hot acid, and rinsed with a few drops of acid back into the beaker.  
The acid was allowed to cool, which typically precipitated the PPO.
By decanting or centrifugation, the acid was extracted into a separate
PE bottle.  The process was then repeated in which a fresh batch
redissolved the residual PPO sample and etched the sample container in
a new beaker, to further extract adsorbed activity.
All of these acid solutions were then collected and assayed.

High purity germanium detector spectrometers at Caltech and UA
were used to assay the radionuclides in the acid.  Two such
detectors exist at each institution for this work.  Each detector
is surrounded by approximately 10 cm of lead and 5 cm of copper to
reduce external backgrounds. It is important to note that a low background
counting facility is essential for this study because radiochemistry
separation methods are employed to remove competing activities. 
The detector performances vary
in relative NaI efficiency from 30 to 85\% at 1332 keV, resolution from
2 to 5 keV, and ambient background from $4\times 10^{-5}$ to $2\times 10^{-3}$ 
$\rm counts/(s\cdot keV)$ at 300 keV.

During the acid counting, one is limited to 10 ppt sensitivity in a 
gram of irradiated material 
for activities tagging uranium and thorium.  The overall spectrometer 
counting rate is dominated by the decay radiation
of $^{82}$Br and $^{24}$Na, originating from sodium and bromine
naturally abundant in the scintillator.  Partial energy
depositions from the Compton scattering  of high energy
gamma radiation typically produce backgrounds
of  1.5 $\rm counts/(s \cdot keV)$ near 300 keV.
 A further radiochemical
step isolates actinides from these sources of backgrounds.  The
acid is poured through a plastic column containing actinide
absorbing resin;  the TRU or Actinide resin of Eichrom Technologies,
Inc. were used.  To reduce the
remaining PPO precipitant in the acid which may clog the column, the acid
was prefiltered with either a column
of inactive resin or a disposable plastic filter.
Ideally, the column selectively absorbs actinide activities while
allowing sodium, bromine, and other light elements to pass through
harmlessly.  In practice, we noted a reduction of greater
than 100 for the  sodium activity, 
and greater than 5 for bromine, for a typical background
in the $^{233}$Pa and $^{239}$Np region of interest of
$4 - 80\times 10^{-3} {\rm counts/(s\cdot keV)}$.  

The 100 ml polyethylene bottle containing the extracted
nitric acid was counted
in a high resolution germanium detector for 12 hours or more, in order
to detect the presence of short-lived nuclei.  The acid counting focused
on $^{42}$K detection.
The column was sealed in a bag and taped to the top of a Germanium
detector for radiation assay; the column's small size ( 1 cm diameter
by 5 cm long) provided for a high efficiency in gamma ray
detection.  The column data provided a sensitive measure of uranium
and thorium related activity.
\begin{figure}[hhh!!!]
\begin{center}
\leavevmode
\mbox{\psfig{file=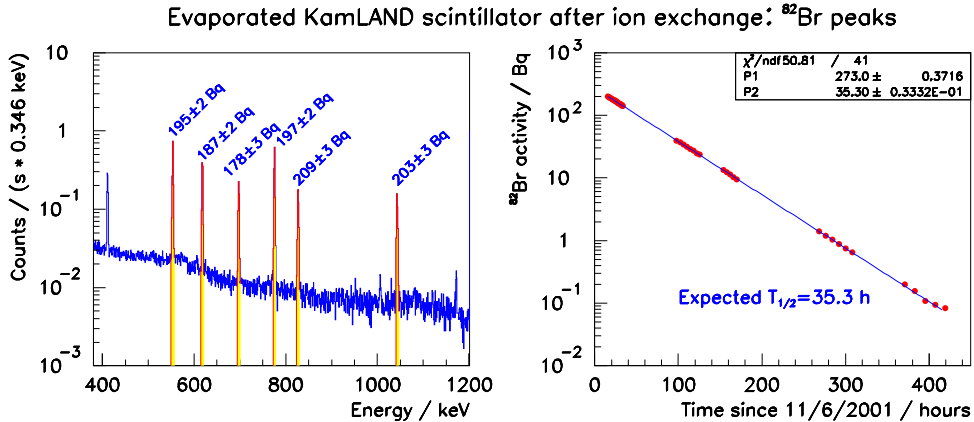,width=15cm,clip=}}
\end{center}
\caption{Part of a typical gamma spectrum (left) and time evolution (right) 
of the determined activities. The parameters p1 and p2 are denoting the activity at time
zero in Bq and the half life in hours, respectively.}
\label{br_fig}
\end{figure}
The detection efficiency of the germanium counters was determined
using nine-activity solutions
from Isotope Products
Laboratories and Northern Analytics 
with gamma activities spanning 88 to 1836 keV calibrated
to 3\%.  
The calibration solution was diluted with acid and then filled into
containers with identical counting geometry and liquid volume as
those used in the neutron activation analysis.  The detector efficiency in
each geometry was determined for each calibrated activity, and 
a six parameter calibration function was fitted to the data.
The absolute uncertainty on the germanium detectors' efficiencies
from this determination
is 6\%.  In practice, repeated measurements of the same sample
showed an activity variance of 10\%, where the spread results from 
imprecision in geometric sample placement around the detector.\\
As an example for the counting procedure Fig.~\ref{br_fig} depicts
the analysis of $^{82}$Br in activated scintillator residue.
The left panel of Fig.~\ref{br_fig} shows part of a typical gamma spectrum 
obtained after counting the ion exchange column. The six peaks are due to the decay of $^{82}$Br,
one of the main interferences. The good energy resolution and large peak to Compton ratio of
the Germanium counter is used to isolate the gamma lines. Partial activities displayed for
every peak show that the counting efficiency correction is consistent. The right panel
shows the temporal development of the weighted average of all partial
Br activities. A least squares fit to an exponential decay 
is used to determine the activity at time zero. This double differential analysis allows for
an unambiguous identification of the various radio-nuclides contained in the sample and
offers a compact way to summarize the numerous peak integrals. For the quantitative
analysis the half life was fixed to the known value. The masses of the elements of 
interest were then inferred from the measured activities using the known neutron flux
and energy distribution. The equations governing this calculation are well 
known~\cite{naa_book} and shall not be repeated here.

A full suite of samples for neutron activation analysis included 
one or more copies of a KamLAND scintillator fluid for analysis,
in addition to a neutron fluence standard, 
one or more spiked scintillator calibration standards,
an analysis reproducibility standard, and a complementary pure water blank
for each of the preceding materials.  Not all of these samples
were available during all irradiations. Procedures describing contamination
controls are included in the Appendix.

\section{Determination of the Efficiency}
Samples of liquid scintillator quantitatively prepared with large
trace element impurities monitored the overall efficiency of the analysis.  
Metallo-organic element standards dissolved in mineral oil were
purchased from Alfa Aesar for potassium, tin, and lanthanum.
For uranium and thorium, no commercial organic standard was available, so
acid sources were extracted into a form soluble in organic
liquids by tri-n-butyl phosphate(TBP) extraction~\cite{tbp}.  For uranium, 
uranyl nitrate dissolved in 5ml 1M HCl was washed with three
lots of 5 ml TBP. 
For thorium,  thorium nitrate crystals were dissolved in a 15 ml solution
of 7M ${\rm Ca(CO_3)_2/2M\; HNO_3}$.  This was washed with three 15 ml lots of TBP.
In each case, the TBP
or commercial standard was dissolved in liquid scintillator and 
diluted to near 10 ppb element impurity.  ICPMS, NAA, and low background
gamma ray assay were all used to calibrate the final element
impurities with a 10\% 
uncertainty.

For spiked sample preparation and radiochemical analysis, 
completely separate but identical laboratories and equipment
were used in order to avoid cross-contamination.  
Only the germanium detectors used for the counting
were common between the calibration and clean samples.
In an identical evaporator
as described for the clean analysis, an approximately one gram aliquot of
standard 
impure scintillator was mixed with a 50 ml KamLAND scintillator and evaporated.
This calibration sample was collected, bottled, irradiated, 
radiochemically extracted and finally counted for trace activities.

Irradiation of scintillator with known element concentrations yielded
information about the element transfer in the analysis process.  
Element concentrations in the standard spiked scintillator were inferred
from the activities measured in the extraction acid and actinide
column.  The values were compared to the known impurity of the
scintillator to derive an analysis efficiency for the entire
process. The results are summarized in Table~\ref{tab:colefficiency}.
\begin{table}[h]
\begin{center}
\begin{tabular}{|l|p{3.25cm}|p{3.25cm}|}
\hline
\multicolumn{3}{|c|}{Amount Recovered in Actinide Column in [\%]}\\
\hline
Method       & $^{232}$Th    & $^{238}$U    \\ \hline
NAA          &  33$\pm$23  & 67$\pm$40  \\ 
\hline
\end{tabular}
\end{center}
\caption{Percent recovery of uranium and thorium
by radiochemical activation analysis with
column extraction (an Eichrom Actinide Column in this case).  These
efficiencies
calibrate the full analysis, including evaporation, irradiation, and
radiochemical procedures.  
}
\label{tab:colefficiency}
\end{table}
Efficiency is clearly demonstrated for all the analyzed elements, though
the efficiencies do not approach 100\%.

The overall efficiency encompasses a variety of retention and
loss mechanisms
including element loss in evaporation and inefficient activity extraction
into acid and into actinide columns. 
These were measured separately in order to better understand
the sources of inefficiencies.
The retention efficiencies are shown  in Tables ~\ref{tab:efficiency}
and ~\ref{tab:columnrecovery}.
 
\begin{table}[h]
\begin{center}
\begin{tabular}{|l|c|c|c|c|c|}
\hline
\multicolumn{6}{|c|}{Average Amount Recovered in Acid in [\%]}\\
\hline
Method &  $^{40}$K      & $^{112}$Sn     & $^{139}$La     & $^{232}$Th     & $^{238}$U \\ \hline
ICPMS  &                &                &                &                & $82\pm 17$ (7) \\
NAA    & $35\pm 20$ (3) & $50\pm 23$ (2) & $52\pm 15$ (3) & $45\pm 34$ (2) & $73\pm 37$ (5) \\
\hline
\end{tabular}
\end{center}
\caption{Averaged percent recovery of elements, compared to the
expected from a standard impurity,
for an acid digestion
of evaporated liquid scintillator samples. The scintillator evaporation
is expected to dominate the losses. The errors given correspond to the
standard deviation observed for repeated measurements. The number of
calibration measurements performed over the course of this study is
given in brackets.
}
\label{tab:efficiency}
\end{table}
The chemical efficiency of the ion exchange process was calibrated separately.
This determination does not need to rely on knowledge of the element
concentration in the spike solution, it can be done by comparing measured
activities before and after ion exchange. To do that
the analyzed activities were separated between the ion exchange column,
residual acid and solid residue of irradiated sample. The sum of these three
partial activities constitute 100 \% of induced activity.
Repeated studies of the radiochemical
extraction showed little change in the activity recovery with
variations in certain aspects of the above procedure, including
mineral acid composition, acid strength, filtration techniques, vial breaking
technique, and acid digestion temperature.
\begin{table}[h]
\begin{center}
\begin{tabular}{|l|c|c|c|c|c|}
\hline
Retention in   &  $^{24}$Na[\%] &  $^{42}$K[\%] & $^{82}$Br[\%] & $^{233}$Pa[\%] & $^{239}$Np[\%]\\ \hline
Column         & 0.01$\pm$0.01  & 0.1$\pm$0.1   & 17.5$\pm$1.1  & 96.3$\pm$9.5   & 84.8$\pm$8.1  \\
Acid           & 85.1$\pm$6.8   & 94.7$\pm$12.6 & 60.5$\pm$3.4  & 0.4$\pm$0.3    & 1.2$\pm$0.3   \\
\hline
\end{tabular}
\end{center}
\caption{Percent recovered activities in Eichrom TRU resin
column and residual acid, after column extraction. The balance of the
activities stays in the scintillator residue.}
\label{tab:columnrecovery}
\end{table}
As an additional constraint a straightforward separation of $^{237}$Np was 
carried out on Eichrom TRU-Spec columns.  
A solution of $^{237}$Np in 8M nitric acid was employed as a tracer.  15 ml of 
tracer solution was allowed to drain through the column and was
followed by consecutive washes of 8M HNO$_3$.  
The first acid wash
of 4 ml was analyzed for $^{237}$Np by liquid scintillation and was found to contain 
only 2.8 percent of
the total $^{237}$Np added to the column.  Analysis of the elutant during loading of the 
$^{237}$Np tracer
sample showed no detectable breakthrough of the tracer.  The tracer was eluted 
from the column with 93 percent recovery using a single 4 ml water wash 
as determined by liquid scintillation. 
A second column was treated in like manner to establish blank values for 
the liquid scintillation detection.  The reproducibility of the $^{237}$Np
recovery was approximately 5 percent for three determinations.

Activation analyses of clean scintillators are corrected for these
measured efficiencies.  Note that the purpose of this work is
to demonstrate trace element purity of materials, rather than
precisely
calibrate the impurities.  The precision in the determination of
the analysis efficiency in the tables is sufficient for this purpose.
It is noted that the full analysis efficiency for $^{232}$Th and
$^{238}$U, as reported in Table~\ref{tab:colefficiency} is consistent with 
the product of the partial efficiencies measured for the sample evaporation
and column retention, given in Tables~\ref{tab:efficiency} 
and~\ref{tab:columnrecovery}.
The control over the element concentrations in the low level organic spike
solutions used to derive these efficiencies contributes to the relatively
large fluctuations observed. From the above data we conclude that trace
impurity concentration can be determined quantitatively within a factor of
two accuracy when applying the methods and procedures discussed
in this paper.

\section{Results}
\begin{table}[hhh!]
\begin{center}
\begin{tabular}{|l|c|c|c|c|}
\hline
Material       & Mass(g) &  $^{40}$K (g/g)      & $^{232}$Th (g/g)              & $^{238}$U (g/g)  \\\hline
PPO Lot 21-634 & 3.43  & $<8.8\times10^{-12}$   & $(3.2\pm1.1) \times 10^{-12}$ & $<2.7\times 10^{-12}$ \\
Evap. PC+PPO   & 0.86  & $<73\times 10^{-12}$   & $<1.5\times 10^{-12}$         & $<2\times 10^{-11}$   \\
LS Sample \#1  & 500.9 & $(5.3\pm0.7)$          & $<15.8\times 10^{-14}$        & $<5.1\times 10^{-14}$ \\
               &       & $\times 10^{-13}$      &                               &                       \\ 
LS Sample \#2  & 96.7  & $(3.1\pm0.6 )$         & $(1.8\pm 0.6)$                & $<1.3\times 10^{-14}$ \\
               &       & $\times 10^{-14}$      & $\times 10^{-14}$             &                       \\
KamLAND Final  & 353.0 & $<2.4\times 10^{-15}$  & $<5.5\times 10^{-15}$& $<8\times 10^{-15}$   \\
Liquid Scint.  &       &  & & \\ 
\hline
\end{tabular}
\end{center}
\caption{Impurity of KamLAND liquid scintillator materials determined with
activation analysis.  For null signals, 90\% confidence limits on the
contamination are shown.}
\label{tab:lsanalysis}
\end{table}

During the application of the activation analysis for KamLAND,
three
representative organic materials were analyzed.  The impurity
concentrations derived from activation analysis are shown
in Table \ref{tab:lsanalysis}.  Different lots of PPO from Packard
BioScience were
screened for purity before they were mixed with the other scintillator
components.  PPO 21-634 is one such lot.  Since the PPO contributes
only 0.2\% of the final scintillator mass, it was required to be less
than $5\times 10^{-12}$ g/g in radionuclides to reach the reactor 
anti-neutrino goals
defined in the introduction.  The PPO was directly irradiated in plastic
containers, without any evaporation.
During final scintillator production for KamLAND,
all of the PPO lots were predissolved in pseudocumene.  The
concentrated PPO in PC at 200 g/liter was screened prior to
final scintillator mixing.  The mixture was evaporated
and irradiated in plastic vials.  

Liquid scintillator sampled from the KamLAND detector was
finally analyzed using these techniques.  The 1000 tons  were filled
into KamLAND over a six months period.  Samples of the scintillator
fluid were taken as it entered the detector, during
the second and fourth month.  These samples
were evaporated and irradiated in quartz.  Their analysis is labeled
``LS Sample \#1 and \#2'' in the table.
At the end of the filling period a sample of final detector fluid
was taken and analyzed.   Each of the liquid scintillator samples
was evaporated and irradiated in pure quartz vials.  In the filling
samples, a potassium impurity was detected, but it's strength varied between
the two samples studied.  The final detector LS had no detectable 
radioimpurity.

The sensitivity
to these trace radionuclides reached below $8\times 10^{-15}$ g/g, and the 
final scintillator material
was proven pure at this level.  These sensitivities were sufficient
to certify the organic materials as suitable for the KamLAND reactor
anti-neutrino experiment.

\begin{figure}[hhh!!!]
\begin{center}
\leavevmode
\mbox{\psfig{file=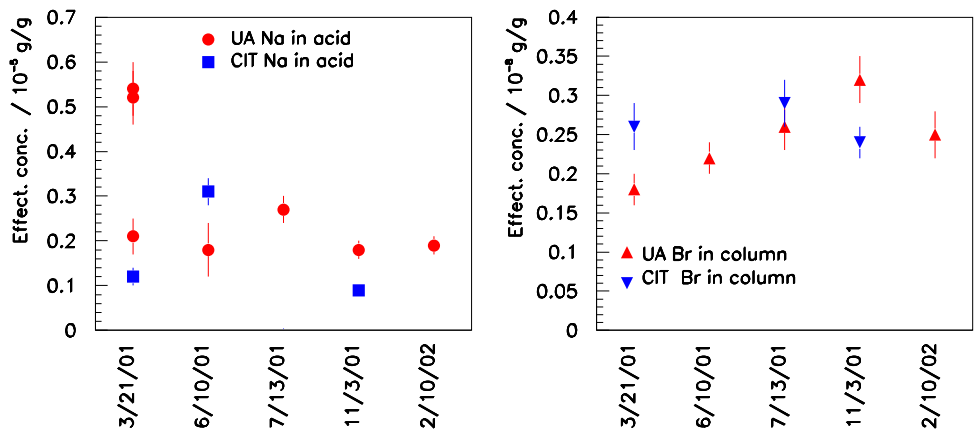,width=15cm,clip=}}
\end{center}
\caption{Analysis consistency as determined through repeated analysis
of PPO batch 21-634. The left part of the figure shows the effective
concentration of Na, as determined from the acid used in radio chemical
post irradiation preparation. The right part shows the effective Br
concentration as determined from counting of the ion exchange columns.}
\label{anal_con}
\end{figure}
Analysis consistency was monitored by repeatedly analyzing
the same material during 5 irradiations.  
The reproducibility and laboratory to laboratory consistency was verified
by repeated analysis of PPO samples of production batch 21-634.
To quantify variations in our sample treatment the effective concentrations
of Na in the acid, discharged from the ion exchange column, and of Br in
the columns was determined. The concentrations are labeled ``effective''
because 100\% retention efficiency in the respective medium has been assumed.
The retention efficiency for Na in acid has been determined to be
(85$\pm$7)\%. The retention of Br in the column is estimated to be about (18$\pm$1)\%.
Na and Br have been chosen because they exhibit the lowest preparation blank
and therefore most robust signature.
Figure~\ref{anal_con} depicts the results of this repeated analysis.
The concentrations derived for Br vary by about $\pm$5\% when
averaged over all activation runs. The average Br concentration
derived at Caltech [$(2.56\pm 0.15)\times 10^{-9}$] and UA
[$(2.31\pm 0.11)\times 10^{-9}$] agree within $1.4\; \sigma$.

\section{Conclusion}

The primordial activities potassium-40, thorium and uranium constitute important sources
of background for low energy low rate experiments. Access to sensitive detection
techniques is crucial in the development of next generation neutrino
oscillation, double beta decay and dark matter experiments. 
The analysis technique described in this article might also be important for the 
development of a next generation ultra
low background counting facility in the planned US National Underground Science
Laboratory.

In this paper we discuss a novel ultra-sensitive neutron activation analysis,
optimized towards the detection of primordial activities in liquid scintillator.
The achieved sensitivity is below $8\times 10^{-15}$ g/g. The described procedure
is applicable at almost any research reactor facility.

{\bf Acknowledgment}\\

This work was supported in part by the US Department of Energy and the US-Japan 
Committee for Cooperation in High Energy Physics. The authors 
would like to thank S.J. Freedman, K.T. Lesko
Y. Kamyshkov, A. Suzuki and J. Webb for help during the critical early
phase of this project.  The authors appreciate the advice and 
assistance of K. Farley,
J. Goreva, L. Hedges, and G. Wasserburg during the ICPMS studies.
We would also like to thank the KamLAND collaboration for their support.

{\bf Appendix}\\

A key concern in high sensitivity analysis is the control
of contamination during the analysis.  A number of details
regarding the measures taken in this work to control contaminations
are included here.
In order to address the level of potential contaminations in sample extraction,
similar containers, both empty and half full with ultrapure water, were
exposed to the air at the liquid access point for one week.  The containers
were filled with hot, concentrated nitric acid for one day, and the acid
was then analyzed by ICPMS.  No uranium or thorium contamination
was detected.  A limit of \mbox{$<9\times 10^{-16}$ g} per liter of fluid maximum 
contamination
from the above collection process was inferred, when properly
extrapolated
to the collection conditions.

Contamination control during evaporation
is another challenge of the technique.  
After evaporation of 600 ml of liquid scintillator 
the typical sample
size of about 800 mg results in a much higher vulnerability towards the introduction of
contaminations. The cleanliness requirements during sample evaporation and handling 
are thus very stringent.

Scintillator evaporation at low temperatures takes very long thus resulting in a prolonged 
exposure to contamination. It was observed that significant loss of PPO occurs if the
scintillator is evaporated at higher temperatures. Vacuum evaporation offered the
best compromise between evaporation speed and PPO retention.
Evaporation of scintillator samples at various temperatures showed that at temperatures
between 63$^{o}$C and 75$^{o}$C 70\% of the PPO is recovered while
the evaporation rate is with about 2 g/h still reasonable. The evaporation rate
scales approximately 
linearly with the temperature and reaches 20 g/h at 95$^{o}$C.
All sample evaporations
were performed in the above temperature range and at a pressure of about 40 torr.

\begin{enumerate}
\item All work was done in class 100 or 500 clean rooms, to avoid sample contamination.
\item The evaporations were performed using Sheldon Model 1415 vacuum ovens.
\item During evaporation vacuum was maintained using an oil free Teflon diaphragm pump.
\end{enumerate}

In order to provide an overall monitor of
contamination during the sample preparation, 
empty quartz or plastic vessels were irradiated in parallel
with samples.  The procedure for
evaporating the scintillator involves transferring the liquid
between several PFA vessels until the PPO residue reaches the
irradiation bottle, so that one is particularly sensitive to
contamination during
this sample handling. This contamination was monitored in practice
by duplicating the procedures with ultrapure water.  Water certified to
purity of $<$5$\times$10$^{-12}$ g/g potassium and $<$5$\times$
10$^{-15}$g/g uranium and thorium was purchased from Tama Chemicals, and
was evaporated in parallel to scintillator samples, using
the same equipment.  At the temperature and pressure used
for scintillator evaporation, the water completely evaporates in
a few hours.  After opening the oven to recover the PPO beaker,
one milliliter of fresh water was added to the blank evaporation beakers
and was poured into quartz irradiation containers.  These were
evaporated to dryness in the same oven and sealed.  The resulting
containers were then irradiated and analyzed in parallel with
the samples, using all the same steps and techniques described
in the text.

\end{document}